\documentclass[a4paper,11pt]{article}

\usepackage[utf8]{inputenc}   % For UTF-8 encoding
\usepackage{amsmath}          % For mathematical symbols and environments
\usepackage{amssymb}          % For additional mathematical symbols
\usepackage{graphicx}         % For including graphics
\usepackage{color}
\usepackage{hyperref}         % For hyperlinks
\usepackage{geometry}         % For adjusting page dimensions
\usepackage{natbib}           % For bibliography
\usepackage{authblk}          % For author affiliations
\usepackage{appendix}
\usepackage{multirow}
\usepackage{tablefootnote}
\usepackage{threeparttable}
\usepackage{longtable} 
\usepackage{booktabs}  % For better tables
\usepackage{amsmath}   % For math symbols
\usepackage{lipsum}    % Optional: For generating text (if needed)

\title{Degeneracy in Accretion Disk Spectra from Naked Singularities and Kerr Black Holes: Application to the AGN MCG–06-30-15}
\author[1,2]{Vishva Patel\thanks{vishvapatelnature@gmail.com}}
\author[3]{Sayantan Bhattacharya\thanks{sayantan.bhattacharya@tifr.res.in}}
\author[3,4]{Sudip Bhattacharyya\thanks{sudip@tifr.res.in}}
\author[2]{Pankaj S. Joshi\thanks{pankaj.joshi@ahduni.edu.in}}

\affil[1]{PDPIAS, Charusat University, Anand 388421, India}
\affil[2]{International Centre for Space and Cosmology, School of Arts and Sciences, Ahmedabad University, Ahmedabad 380009, India}
\affil[3]{Department of Astronomy and Astrophysics, Tata Institute of Fundamental Research, Colaba, Mumbai 400005, India}
\affil[4]{MIT Kavli Institute for Astrophysics and Space Research, 
Massachusetts Institute of Technology, Cambridge, MA, 02139, USA}

\date{}

\begin{document}

\maketitle

\begin{abstract}
Theoretical studies suggest that gravitational collapse can form either a black hole or a visible (naked) singularity. Identifying observational signatures that distinguish these two types of collapsed objects is a holy grail of physics. Here, we examine whether relativistic accretion disk spectra can provide such a test. We construct an additive table model for a thin accretion disk in the Joshi–Malafarina–Narayan (JMN-1) naked singularity geometry matched to a Schwarzschild exterior and fit it to {\it NuSTAR} X-ray data from the AGN MCG–06-30-15. Our results are compared with standard Kerr and Schwarzschild black hole models. We also include the relativistic reflection spectral component \texttt{relxill}. Despite their different underlying geometries, the spinning (Kerr) black hole and the non-spinning JMN-1 naked singularity provide similar spectral fits, which are significantly better fits than the Schwarzschild black hole. This degeneracy between the naked singularity and the Kerr black hole could lead to incorrect spin measurements of collapsed objects using disk spectra. The degeneracy could be broken with an independent spin measurement, which could also help identify a naked singularity. Our results could also have a role in different spin distributions of collapsed objects measured from gravitational-wave sources and X-ray binaries.
\end{abstract}

% Main Content
\twocolumn
\section{Introduction}
\label{sec1}

The nature of the supermassive collapsed objects at the center of galaxies remains one of the most intriguing puzzles in modern astrophysics and gravitation. Although the black hole paradigm, particularly the Kerr solution of general relativity, is the prevailing model to describe these objects, alternative possibilities such as naked singularities have not been definitively ruled out either observationally or theoretically. The cosmic censorship conjecture (CCC), first proposed by Penrose \cite{Penrose:1964wq}, posits that singularities arising from gravitational collapse must always be hidden within an event horizon, ensuring that no information can escape from these regions. However, various exact solutions and numerical simulations challenge this view, suggesting that under certain initial conditions, a gravitational collapse can result in a spacetime with a visible singularity, termed a naked singularity \cite{Joshi:1993zg, Deshingkar:1998ge,Ortiz:2011jw,Debnath:2003pq,Giacomazzo:2011cv,Ori:1989ps, Goswami:2004ne, Joshi:2011zm, Mosani:2020mro, Mosani:2021byw}. This makes the cosmic censorship conjecture still an open, challenging, and unresolved mystery in the field of gravitational physics. \\

In particular, the class of JMN spacetimes offers a compelling model for naked singularities formed via the collapse of matter with tangential pressure gradients. These solutions are regular everywhere except at the central singularity and admit smooth matching with a Schwarzschild exterior at a finite radius. The JMN-1 solution is especially interesting due to its potential astrophysical implications \cite{Joshi:2013dva,Bambhaniya:2019pbr,Dey:2020haf,Shaikh:2019hbm,Trivedi:2025ukh,Tahelyani:2022uxw,Guo:2020tgv,Shaikh:2018lcc,Broderick:2024vjp,Chakraborty:2024jma,Kalsariya:2024qyp,Joshi:2024gog,Azreg-Ainou:2023izc,Saurabh:2023otl,Pal:2022cxb,Acharya:2023vlv}.  Since both black holes (BHs) and naked singularities (NaSs) can mimic similar gravitational potentials at large distances, differentiating them requires analyzing phenomena that occur in their strong-field regimes, particularly where accretion disk physics becomes crucial. Moreover, recent studies have shown that several observable phenomena including orbital precession and the shadow cast by compact objects can exhibit overlapping signatures in both black holes and naked singularities \cite{Bambhaniya:2022xbz}. Recently, the Event Horizon Telescope (EHT) collaboration has released high resolution images of the shadow of compact objects in Sgr A* \cite{EventHorizonTelescope:2022xqj}, sparking renewed interest in alternative models. They analyzed the shadow morphology of various compact object models. They concluded that the JMN-1 naked singularity, despite lacking an event horizon, can closely resemble the shadow of a Schwarzschild black hole. This suggests that JMN-1 can act as a viable black hole mimicker, making it difficult to distinguish the true nature of the compact object based solely on shadow observations. \\

Another key feature is the accretion disk, one of the most luminous and informative astrophysical structures, powered by the conversion of gravitational potential energy into radiation as matter spirals inward toward the central object. The emitted X-ray spectra from these disks encode essential information about the geometry of the underlying spacetime \cite{Liu:2018bfx,Bambi:2021hxv,Bambi:2022iuz,Dovciak:2003jym,Psaltis:2008bb,Reynolds:2011sba,Johannsen:2012ng,Miller:2012zj,Bambi:2012tg,Bambi:2013qj,Bambi:2013hza,Jiang:2014loa,Johannsen:2014arp,Riaz:2022rlx,Kurmanov:2025uwq}. By studying the spectral and flux properties of such disks, it is possible to test general relativity in its most extreme environments and distinguish between different spacetimes. \\

One powerful approach to perform this comparison is by constructing models of accretion disk spectra for both BHs and NaSs and fitting these against observational data. The Novikov-Thorne thin-disk model \cite{Novikov:1973kta,Page:1974he}, when extended to general spherically symmetric static geometries, enables such calculations by allowing for the inclusion of physically motivated metrics, such as the JMN-1 naked singularity spacetime. Moreover, when spacetime features a matching boundary between interior and exterior solutions, care must be taken while accounting for the physical quantities such as angular velocity and energy flux. Analytical expressions for particle motion and energy dissipation derived from the metric are essential for constructing realistic emission models in such mixed geometries \cite{Tahelyani:2022uxw,Guo:2020tgv}. \\

Among the ideal candidates for such a study is the Narrow Line Seyfert 1 galaxy MCG-06-30-15, located at a redshift of $z=0.00775$. This source has been extensively analyzed due to its pronounced soft X-ray excess, strong reflection features, and rapid variability in both soft and hard X-ray bands, indicating emission from regions very close to the central compact object \cite{Tripathi:2020cje,Jiang:2022sqv,Tripathi:2017uif,Kammoun:2017wcq,Lira:2015wya,Emmanoulopoulos:2011ur,Reynolds:1995mt,Iwasawa:1996uh,Orr:1997mf,Weaver:1998ad,Nowak:1999dj,Fabian:2002gj,Zycki:2010jd,Noda:2011my,Parker:2013cia,Marinucci:2014ita,Gupta:2018soy}. The central engine is estimated to have a mass of approximately $1.6 \times 10^{6} M_{\odot}$, inferred from reverberation mapping and spectral variability studies. Importantly, recent \textit{XRISM} spectral modeling of MCG-06-30-15 also suggests a high spin parameter for the central object, allowing the accretion disk to extend deep into the gravitational potential well \cite{brenneman2025sharper}. This makes the source an excellent testbed for strong-field gravity, as such deep-disk emission is highly sensitive to the near-horizon spacetime geometry. In this context, the JMN-1 naked singularity offers an intriguing alternative. Despite being non-spinning, it permits stable circular orbits all the way to the central singularity, effectively mimicking the deep accretion structure of a high-spin Kerr black hole. This unique overlap in disk properties allows for a meaningful comparison between spinning black hole models and non-spinning horizonless alternatives, such as JMN-1,  using real astrophysical data. These characteristics make MCG-06-30-15 a compelling laboratory for testing deviations from the Kerr paradigm.\\

Motivated by this, we aim to systematically investigate whether naked singularity spacetimes can serve as viable alternatives to black holes in modeling the high-energy emission from accretion disks, and whether accretion disk spectra can help distinguish between these two classes of compact objects. We construct an additive table model based on the JMN-1 solution matched to a Schwarzschild exterior and use it to fit the \textit{NuSTAR} spectra of MCG-06-30-15. We also compare the results with standard models like \texttt{$relxill$} to assess relative fit quality.\\

The outline of the paper is as follows. In Sec. \ref{sec2}, we present the spacetime geometries of the Schwarzschild black hole and the JMN-1 naked singularity. In Sec. \ref{sec3}, we analyze the dynamics of test particles and derive key physical quantities such as specific energy, angular momentum, and angular velocity in static, spherically symmetric spacetimes. We also obtain the flux and luminosity profiles of thin accretion disks, including scenarios with matching hypersurfaces. In Sec. \ref{sectablemodel}, we describe the construction of an additive table model based on the JMN-1 metric and the corresponding luminosity expressions. Sec. \ref{sec5} is devoted to the spectral analysis of the active galactic nucleus MCG-06-30-15 using data from \textit{NuSTAR}. We perform comparative fitting using both standard relativistic disk models and our naked singularity-based table model. Finally, in Sec. \ref{discussion}, we present our conclusions. Unless otherwise stated, throughout this work, we adopt natural units such that \( c = h = G = \sigma = k = 1 \), where \( c \) is the speed of light, \( h \) is Planck’s constant, \( G \) is Newton’s gravitational constant, \( \sigma \) is the Stefan-Boltzmann constant, and \( k \) is Boltzmann’s constant. We use the metric signature \((- + + +)\). 

\section{\label{sec2}Geometric Structures of Black Holes and Naked Singularities}

\subsection{\label{subsec1}Schwarzschild Black Hole}

In general relativity, the Schwarzschild black hole represents a static, spherically symmetric vacuum solution characterized solely by mass \( M \). Its line element in Schwarzschild coordinates \((t, r, \theta, \phi)\) is given by:
\begin{multline}
    ds^2 = -\left(1 - \frac{2M}{r}\right) dt^2 + \left(1 - \frac{2M}{r}\right)^{-1} dr^2 \\ + r^2 d\Omega^2,
\end{multline}
where \( d\Omega^2 = d\theta^2 + \sin^2\theta \, d\phi^2 \) is the metric on the 2-sphere. The spacetime possesses an event horizon at \( r = 2M \), and no event below this critical radius can send a signal to any external observer. 

\subsection{\label{subsec2}Joshi-Malafarina-Narayan (JMN-1) Naked Singularity}
The Joshi-Malafarina-Narayan (JMN-1) naked singularity spacetime arises from a spherically symmetric collapse of matter with anisotropic pressure and lacks an event horizon \cite{Joshi:2011zm}. The JMN-1 interior solution is matched to a Schwarzschild exterior at a boundary radius \( R_b \) ($M_{0}\,R_{b} = 2M$), and its metric is given by:

\begin{multline}
    ds^2 = - (1 - M_{0}) \left( \frac{r}{R_b} \right)^{\frac{M_{0}}{1 - M_{0}}} dt^2 \\
    + \left(1 - M_{0}\right)^{-1} dr^2 + r^2 d\Omega^2.
\end{multline}

where \( M_0 \in (0,1) \) is a dimensionless parameter related to the compactness of the object.

\section{Particle Dynamics and Accretion Disk Luminosity}
\label{sec3}
We consider the motion of a massive test particle in a general static, spherically symmetric spacetime with metric:
\begin{equation}
    ds^2 = -A(r)\,dt^2 + B(r)\,dr^2 + r^2(d\theta^2 + \sin^2\theta\,d\phi^2).
\end{equation}

In such a background, timelike particles follow geodesics determined by the spacetime geometry. For circular orbits (\(\dot{r} = 0\)), the conserved quantities are: the total energy per unit mass measured at infinity (E), angular momentum per unit mass about the symmetry axis (L), the rate of change of the azimuthal coordinate with respect to coordinate time known as angular velocity $(\Omega)$.\\

The expressions for these quantities are derived from the Euler-Lagrange equations and normalization of the four-velocity:

\begin{align*}
E &= \sqrt{\frac{2 A(r)^2}{2 A(r) - r A'(r)}}, \\
L &= \sqrt{\frac{r^3 A'(r)}{2 A(r) - r A'(r)}}, \\
\Omega &= \sqrt{\frac{A'(r)}{2r}}, \\
g &= -A(r) r^2 B(r).
\end{align*}

Here, \(A'(r)\) denotes the derivative of the metric function \(A(r)\) with respect to \(r\). The negative metric determinant \(g\) is important for computing invariant volume elements and physical fluxes. These expressions are critical for modeling accretion disks, where particles follow nearly circular, stable orbits and radiate as they lose energy \cite{Tahelyani:2022uxw,Guo:2020tgv}.

 \begin{figure*}[h!]
\centering
\begin{minipage}{0.49\textwidth}
\centering
\includegraphics[width=0.81\textwidth]{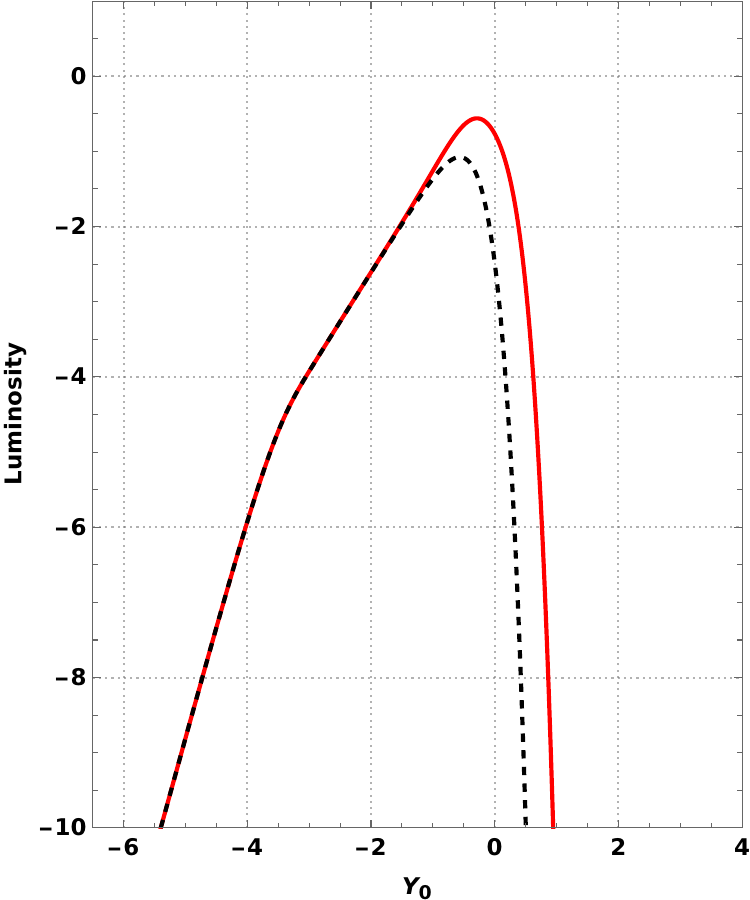}
 \caption{$M_{0}$ = 0.30, $R_{b}$ = 6.67, $R_{in}$ = $10^{-1}$.}
 \end{minipage}
\begin{minipage}{0.49\textwidth}
\centering
\includegraphics[width=0.81\textwidth]{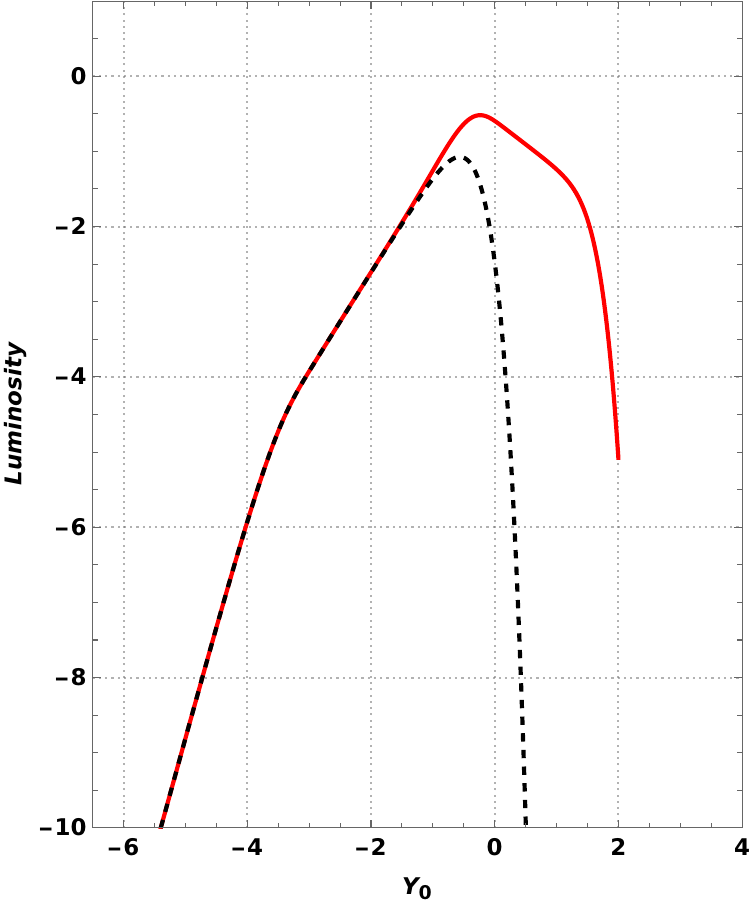}
\caption{$M_{0}$ = 0.30, $R_{b}$ = 6.67, $R_{in}$ = $10^{-5}$.}
\end{minipage}
\begin{minipage}{0.49\textwidth}
    \centering
    \includegraphics[width=0.81\textwidth]{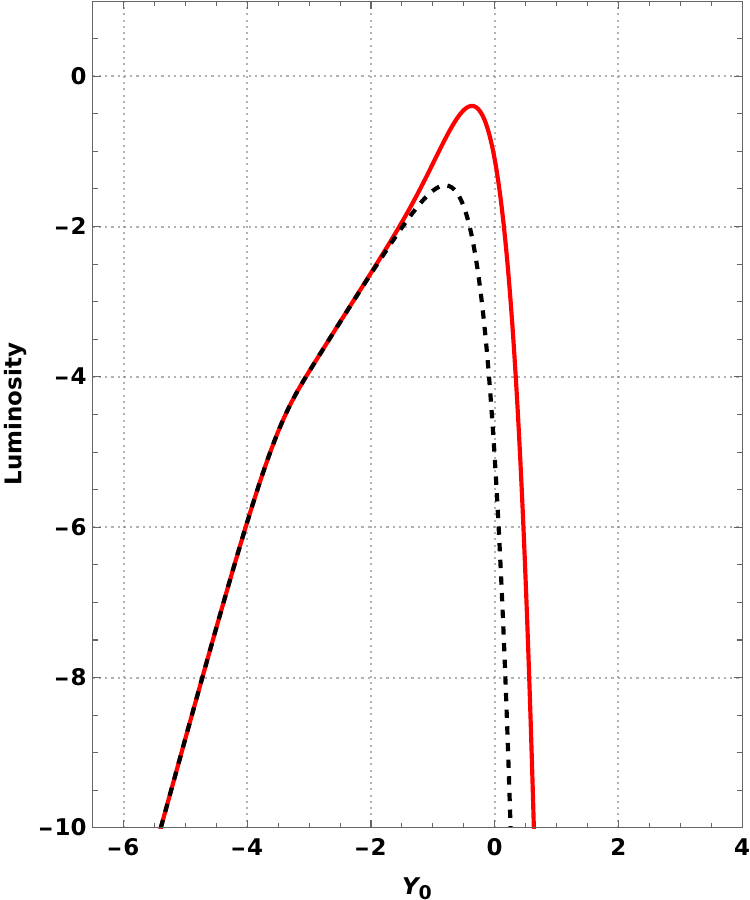}
   \caption{$M_{0}$ = 0.50, $R_{b}$ = 4.00, $R_{in}$ = $10^{-1}$.}
\end{minipage}
\begin{minipage}{0.49\textwidth}
    \centering
    \includegraphics[width=0.81\textwidth]{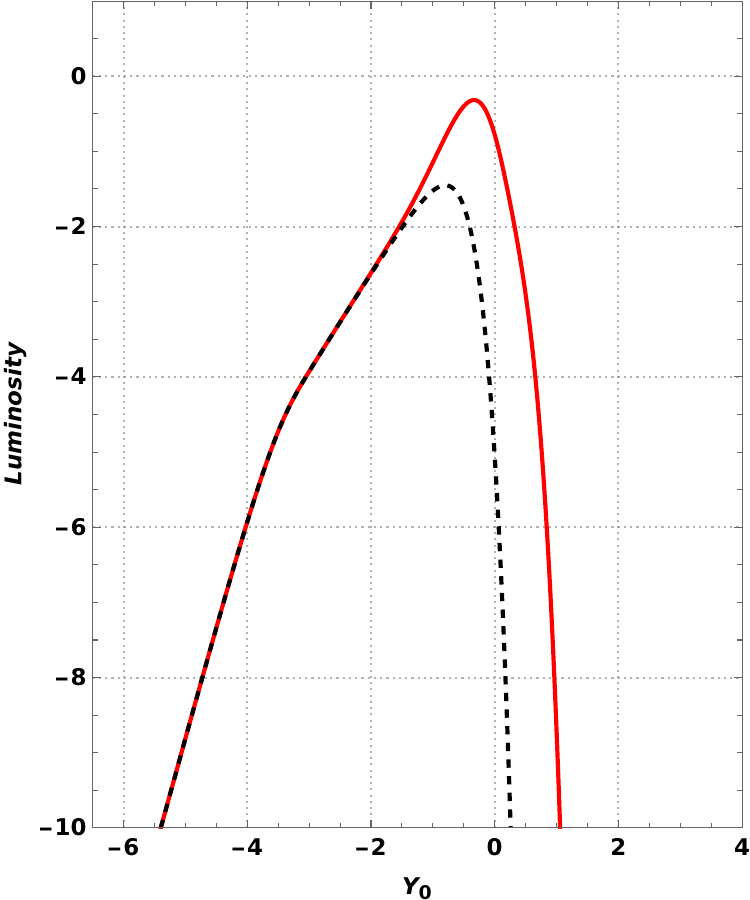}
    \caption{$M_{0}$ = 0.50, $R_{b}$ = 4.00, $R_{in}$ = $10^{-5}$.}
\end{minipage}
\caption{The red line and the black dotted line represent the luminosity spectra for JMN-1 naked singularity and Schwarzschild black hole, respectively. Here, $Y_{0}$ represents the photon energy as given by Eq.\,(\ref{Y0}) }
    \label{fig:luminosity}
\end{figure*}

\subsection{Spacetimes with Matching Boundaries}

In scenarios where the spacetime consists of two distinct regions matched at a boundary radius \( R_b \), such as an interior naked singularity solution and an exterior Schwarzschild solution, the energy flux expression must be modified to account for the discontinuity at \( R_b \). The total energy flux \( \mathcal{F}_{\text{total}}(r) \) is then given by:

\begin{multline}
    \mathcal{F}_{\text{total}}(r) = - \frac{\dot{M}}{4\pi \sqrt{-g_2}} \left( \frac{\Phi_2(R_b)}{\Phi_1(R_b)} \right) \frac{d\Omega_2/dr}{(E_2 - \Omega_2 L_2)^2} \\
    \times \left[ \int_{R_{\text{in}}}^{R_b} (E_1 - \Omega_1 L_1) \frac{dL_1}{dr} dr + \right. \\
    \left. \int_{R_b}^{r} (E_2 - \Omega_2 L_2) \frac{dL_2}{dr} dr \right].
\end{multline}
  
where, subscripts 1 and 2 denote quantities in the interior (JMN-1) and exterior regions (SCH), respectively, $\dot{M}$ is the mass accretion rate, $R_{\text{in}} $ is the radius of the innermost stable circular orbit. This expression accounts for the energy transported by viscous stresses within the disk and radiated away from its surface. The notation $\Phi(r)$ is, $$ \Phi (r) = \frac{(E - \Omega L)^2}{-d\Omega/dr}. $$ 

This expression ensures the continuity of physical quantities across the boundary and accurately captures the influence of the interior geometry on the disk's emission properties.\\

\subsection{ Luminosity Calculation}

The total luminosity \( L \) emitted by the accretion disk is obtained by integrating the energy flux over the disk's surface:
\begin{multline}
    \nu  \mathcal{L}_{\nu,\infty}  timing= \frac{15}{\pi^{4}} \int_{R_{in}}^{\infty} \left(\frac{d\,\mathcal{L_{\infty}}}{d\,ln\,r}\right) \times \\
    \frac{(1+ z)^{4} (h\,\nu/ k T_{*})^{4} / \mathcal{F} }{exp[(1+z)(h\,\nu/ k T_{*}) / \mathcal{F}^{1/4} ] - 1} d\,ln\,r, 
\end{multline}

where the differential luminosity is given by,
\begin{equation}
    \frac{d\,\mathcal{L_{\infty}}}{d\,ln\,r} = 4\pi r \sqrt{-g}\,E\,\mathcal{F}.
\end{equation}

The expressions derived above highlight the dependence of the energy flux and luminosity on the spacetime geometry through the metric functions \( A(r) \) and \( B(r) \). In particular, the presence of a matching boundary introduces additional complexity, as the disk emission properties are influenced by both the interior and exterior regions. Fig.\,(\ref{fig:luminosity}) shows the change in luminosity with respect to $Y_{0}$. Where, $Y_{0}$ is,
\begin{equation}
    Y_{0} = \frac{h \nu}{k T_{*}},
    \label{Y0}
\end{equation}
here, h is Planck’s constant, $\nu$ is the frequency of the photon, k is Boltzmann’s constant and $T_{*}$ is the characteristic temperature of the disk. These constants are set to unity.

\begin{table*}
\centering
\begin{threeparttable}
\caption{\label{tab}
Best-fit spectral parameters and statistical results for different accretion disk models applied to the NuSTAR ObsID 60001047002 X-ray spectrum in the 3--79 keV band using \texttt{tbabs*(diskbb/kerbb/NSJMN + relxill)}. Errors correspond to 90\% confidence intervals.}

\begin{tabular}{| l | c | c | c | c |}
\hline
Model & Parameter (unit) & JMN & Kerr & SCH \\
\hline

\texttt{tbabs} & $n_{\rm H}$ ($10^{22}\ \text{cm}^{-2}$)
& $0.13^{+0.63}_{-0.63}$
& $0.82^{+0.79}_{-0.79}$
& $0.62^{+0.33}_{-0.33}$ \\

\hline

\multirow{3}{*}{\texttt{Disk}}
& $M_0$\tnote{a}
& 0.3
& $6.10\times10^{-2}\,^{+4.97}_{-4.97}$\tnote{b}
& $0.215^{+408}_{-408}$\tnote{c} \\

& $R_b$\tnote{d}
& 6.66
& ---
& --- \\

& $R_{\max}$\tnote{e}
& 1000
& 1000
& 1000 \\

\hline

\multirow{6}{*}{\texttt{relxill}}
& Spin $a$
& 0
& $0.88^{+0.21}_{-0.21}$
& 0 \\

& Inclination (deg)
& $35.7^{+2.0}_{-2.0}$
& $40.7^{+2.5}_{-2.5}$
& $30.0^{+2.2}_{-2.2}$ \\

& Photon index $\Gamma$
& $2.59^{+0.03}_{-0.03}$
& $2.59^{+0.05}_{-0.05}$
& $2.31^{+0.05}_{-0.05}$ \\

& $\log \xi$
& $1.69^{+0.27}_{-0.27}$
& $1.70^{+0.38}_{-0.38}$
& $2.90^{+0.11}_{-0.11}$ \\

& $A_{\rm Fe}$
& $0.60^{+0.09}_{-0.09}$
& $0.75^{+0.12}_{-0.12}$
& $0.50^{+0.19}_{-0.19}$ \\

& Reflection fraction
& $6.18^{+1.22}_{-1.22}$
& $8.00^{+1.30}_{-1.30}$
& $4.23^{+1.31}_{-1.31}$ \\

\hline

\multirow{5}{*}{Statistics}
& $\chi^2$ & 481.63 & 484.02 & 540.26 \\
& dof & 342 & 343 & 344 \\
& $\chi^2/\text{dof}$ & 1.41 & 1.41 & 1.57 \\
& AIC & 497.63 & 498.02 & 552.26 \\
& BIC & 528.51 & 524.41 & 575.46 \\

\hline
\end{tabular}

\begin{tablenotes}[flushleft]
\footnotesize
\item[a] $M_0$ is the compactness parameter of the JMN-1 spacetime.
\item[b] Mass parameter when using the \texttt{kerrbb} model.
\item[c] Inner disk temperature $T_{\rm in}$ (keV) when using the \texttt{diskbb} model.
\item[d] Matching radius between interior JMN-1 and exterior Schwarzschild geometry ($M_0 R_b = 2M$).
\item[e] Outer radius of the accretion disk.
\end{tablenotes}

\end{threeparttable}
\end{table*}

\section{Construction of the Table Model}
\label{sectablemodel}

To compare the spectral signatures of black holes and naked singularities, we constructed a custom additive table model in the FITS format for use within XSPEC \cite{arnaud2003x}. The model is based on a composite spacetime geometry consisting of a naked singularity core matched to a Schwarzschild exterior. This setup reflects the JMN-1 metric in the interior ($r < R_b$) and the Schwarzschild solution outside ($r > R_b$), with the matching radius fixed at $R_b = 2M/M_{0}$ for a mass parameter $M = 1$ and a density parameter $M_0 = 0.3$.

The spectral model calculates the total luminosity of the accretion disk by integrating the energy dissipation rates over radial shells in both regions. Symbolic expressions for the disk luminosity per unit area are defined using \texttt{sympy}, while numerical evaluation is performed using \texttt{scipy}'s integration routines. The integration domain is split across the matching radius to accommodate different expressions for the interior (JMN-1 naked singularity) and exterior (Schwarzschild black hole) regions.\\

A grid of maximum radial extents ($r_{\mathrm{max}}$) ranging from $R_b$ to $10^5 M$ was sampled to generate model spectra. The photon energy array spans the range $10^{-10}$ to $10^3$ in geometrized units (converted to keV via $\log_{10}(E/M) + 13.792$), and the resulting luminosity spectra are converted to physical units ($\log_{10}(L/M^2) + 61.547$) for being suitable to XSPEC.

\section{Observational Analysis}
\label{sec5}
In this section, we describe the methodology adopted for the analysis of X-ray spectral data obtained from the source MCG–06-30-15 using the \textit{NuSTAR} observatory \cite{harrison2013nuclear}. The aim is to investigate the viability of naked singularity models as alternatives to standard black hole solutions by comparing theoretical predictions of disk spectra with observed data. We specifically used a custom additive table model derived from the Joshi-Malafarina-Narayan (JMN-1) spacetime and compared it with standard relativistic disk models.\\

Each spectrum is written to a FITS file using the \texttt{XspecTableModelAdditive} class, which handles interpolation and XSPEC-compatible formatting. The resulting table model, named \texttt{NaSJMN.fits}, can be loaded into XSPEC as an additive model and used in conjunction with standard absorption and Gaussian components to fit observational data.

 \begin{figure*}[h!]
\centering
\begin{minipage}{0.49\textwidth}
\centering
{\includegraphics[width=0.81\textwidth]{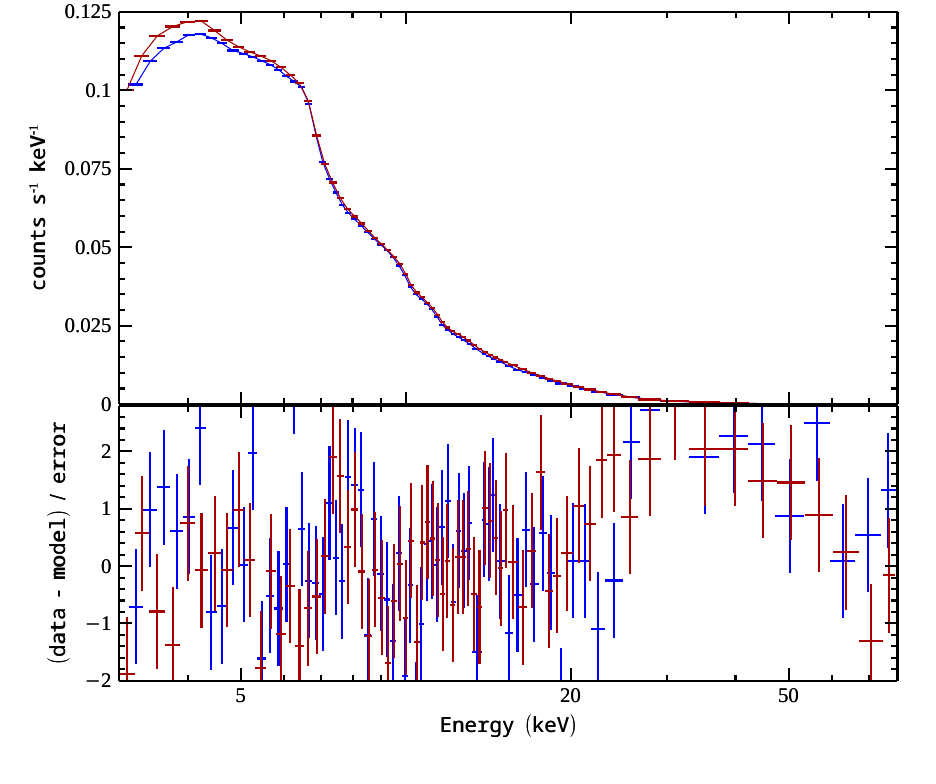}\label{jmndata}}
 \caption{The JMN-1 naked singularity model}
 \end{minipage}
\begin{minipage}{0.49\textwidth}
\centering
{\includegraphics[width=0.81\textwidth]{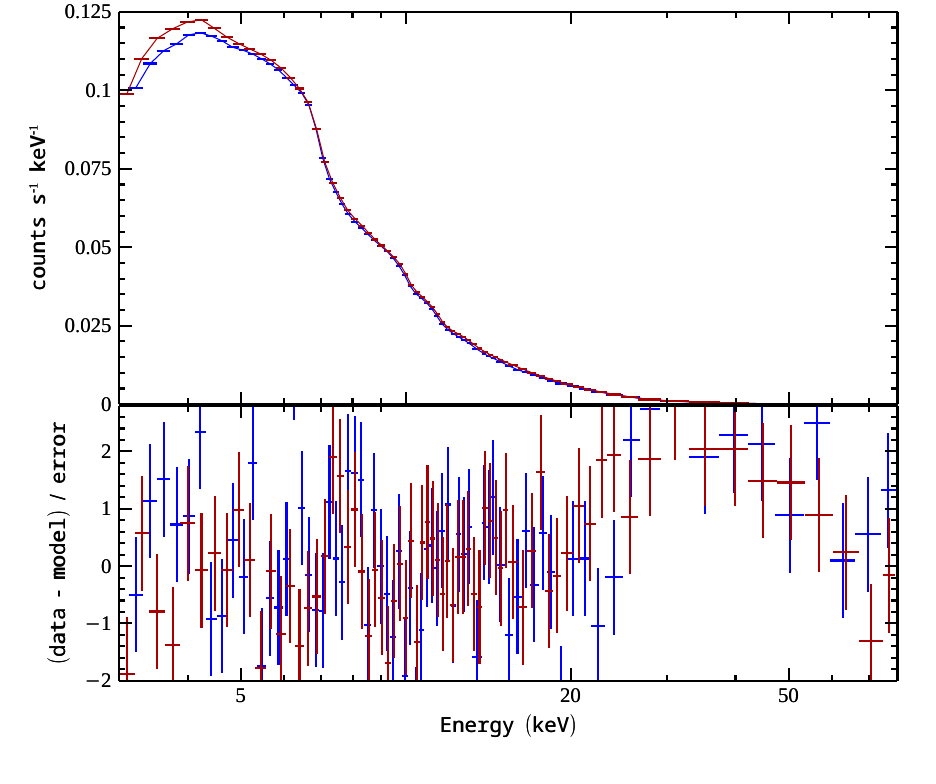}\label{kerrdata}}
\caption{The Kerr black hole model}
\end{minipage}
\begin{minipage}{0.49\textwidth}
    \centering
    {\includegraphics[width=0.81\textwidth]{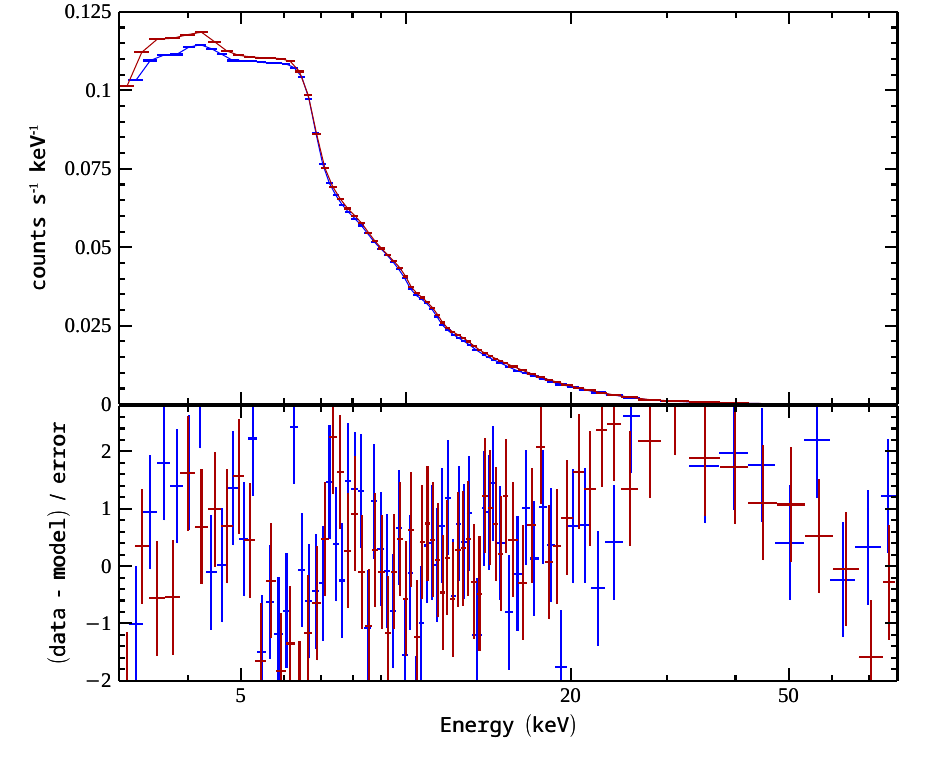}\label{schldata}}
   \caption{The Schwarzschild black hole model}
\end{minipage}

\caption{Spectral fitting of the NuSTAR observation $(\mathrm{ObsID}:\ \texttt{60001047002})$ of the AGN source MCG-06-30-15 using different spacetime models: (a) the JMN-1 naked singularity model, (b) the Kerr black hole model, and (c) the Schwarzschild black hole model. The red and blue data points correspond to the FPMA and FPMB spectrum respectively. The upper panels show the observed counts per keV along with the best-fit model, while the lower panels display the residuals $((\mathrm{data}-\mathrm{model})/\mathrm{error})$.}
    \label{figdata}
\end{figure*}

\subsection{The Source: MCG-06-30-15}

MCG-06-30-15 is a Narrow Line Seyfert 1 (NLS1) galaxy with a redshift of \( z = 0.00775 \), widely studied for its distinctive soft X-ray excess and rapid variability. These features suggest that its X-ray emission arises from very small radii, near the central compact object, where relativistic effects become prominent. The central source is estimated to have a mass of \( \sim 1.6 \times 10^6~M_\odot \), based on reverberation mapping and timing analysis. Given its deep-disk emission characteristics, this AGN is particularly well suited for probing strong gravity regimes.\\

In this analysis, we utilized high-quality broadband observations from \textit{NuSTAR}, which are ideal for constraining both thermal disk emission and relativistic reflection features across the 3--79 keV range. This allows us to perform a robust comparison between standard black hole models and alternative geometries such as the JMN-1 naked singularity.

\subsection{Data Reduction}

We analyzed archival \textit{NuSTAR} observations of MCG-06-30-15 (ObsID: \texttt{60001047002}), downloaded from the HEASARC archive. Data reduction was performed using the \texttt{NuSTARDAS} pipeline, HEASoft (version 6.32), along with the latest calibration database (CALDB). The cleaned event files were generated separately for both focal plane modules: FPMA and FPMB.\\

The reduction pipeline included standard filtering criteria. We applied the \texttt{nupipeline} task to generate cleaned event files, accounting for standard screening of South Atlantic Anomaly (SAA) passages, Earth occultation, and high background periods. For spectral extraction, we used circular source regions with radii of approximately 40" for both FPMA and FPMB, centered on the source coordinates. Background spectra were extracted from nearby source-free circular regions with radii of about 80" for both FPMA and FPMB. These regions were carefully chosen to avoid contamination from the source and other nearby objects. The spectral products were generated using the \texttt{nuproducts}, producing source and background spectra along with the corresponding response matrix files (RMFs) and ancillary response files (ARFs).\\

The extracted spectra were grouped to a minimum of 20 counts per bin using \texttt{grppha}, allowing the use of $\chi^2$ statistics in the fitting process. All spectral modeling was performed using \texttt{XSPEC} (version 12.12.0). Both the standard \texttt{relxill} model and the custom JMN-1-based table model were used for comparative analysis.

\subsection{Spectral Modeling in XSPEC}

Spectral fits were carried out over the 3–79 keV band using the XSPEC analysis package. Interstellar absorption was modeled with \texttt{tbabs}, fixing the hydrogen column density to the Galactic value along the line of sight ($N_H = 0.041 \times 10^{22}$ cm$^{-2}$). A cross-calibration constant was included to account for normalization differences between FPMA and FPMB. The initial fits adopted the form
\[
\texttt{ tbabs * model},
\]

where \texttt{model} corresponds either to the relativistic reflection model \texttt{relxill}\footnote{\texttt{relxill} is a relativistic reflection model developed by \href{https://www.sternwarte.uni-erlangen.de/~dauser/research/relxill/}{J. García and T. Dauser.}} or to a tabulated spectrum based on the JMN-1 naked singularity spacetime. Here, \texttt{relxill\textsubscript{Kerr}} refers to \texttt{relxill} model with non-zero spin parameter for Kerr black hole and \texttt{relxill\textsubscript{SCH}} refers to \texttt{relxill} model with zero spin parameter for Schwarzschild black hole. The additive table model (\texttt{atable\{NaSJMN.fits\}}) contains precomputed spectra covering a range of physically relevant parameters, such as the inner disk radius $R_{\mathrm{in}}$, compactness parameter $M_0$, and boundary radius $R_b$. These spectra were generated on a parameter grid motivated by theoretical considerations, using the general relativistic luminosity formalism described earlier in the paper. The table presenting the best-fitting parameters for the initial fits is given in Table-\ref{tab}. We did not include \textit{diskbb} or \textit{kerrbb} in the final analysis, as standard thermal disk models are not appropriate for AGN X-ray spectra, which are dominated by coronal emission and relativistic reflection. When tested (see Table~\ref{tab}), these models give statistically poorer fits and physically inappropriate parameter values (e.g., unrealistic $T_{\rm in}$ and mass estimates), and were therefore excluded.

\subsection{Fitting Procedure}

For each model, we performed simultaneous fitting of the FPMA and FPMB spectra, allowing the relative normalization constants to vary. Parameters such as inner radius $R_{in}$, inclination angle $i$, and black hole spin $a_*$ (when applicable) were left free to vary within reasonable astrophysical limits. 

Fits were evaluated with the $\chi^2 $ statistic. To compare the relative performance of competing models, we also computed the Akaike Information Criterion (AIC) and the Bayesian Information Criterion (BIC).
\begin{equation}
    \mathrm{AIC} = 2k - 2 \ln(L)
\end{equation}
\begin{equation}
    \mathrm{BIC} = k \ln(n) - 2 \ln(L)
\end{equation}

where $k$ is the number of free parameters, $n$ is the number of data points, and $\mathcal{L}$ is the maximum likelihood. Lower AIC or BIC values indicate a better trade-off between model fit and complexity. In addition to statistical performance, we assessed the physical plausibility of the best-fit parameters to guide model selection. Fig.\,\ref{figdata} presents the spectral fitting of the NuSTAR observation $(ObsID: \texttt{60001047002})$ for the active galactic nucleus (AGN) MCG–06-30-15, analyzed using two different spacetime models: Fig.\,\ref{jmndata}, Fig.\,\ref{kerrdata} and Fig.\,\ref{schldata} represents the JMN-1 naked singularity, the Kerr black hole and Schwarzschild black hole model respectively. The red and blue data points correspond to the spectra obtained from the FPMA and FPMB detectors, respectively. In both panels, the upper plots display the observed counts per keV overlaid with the best-fit model for each case, while the lower plots show the residuals, defined as $((data - model)/error)$, highlighting the deviations between the observed data and the model predictions for each detector.

\section{Discussion}
\label{discussion}    
    In this study, we first discuss the geometric structure of the Schwarzschild black hole and the JMN-1 naked singularity in Sec. \ref{sec2}. This was followed by an analysis of the accretion disk luminosity in both geometries. Figure~\ref{fig:luminosity} shows that the JMN-1 spacetime gives a brighter spectrum at high frequencies than Schwarzschild, especially from the inner disk. We then investigated whether relativistic accretion disk spectra from the AGN source MCG–06-30-15 can be used to differentiate between a black hole and a naked singularity as the central compact object. Using a relativistic reflection framework, we modeled the disk emission for the JMN-1 naked singularity, Kerr, and Schwarzschild spacetimes. Fits to the NuSTAR data allowed us to compare them. We summarize our main findings as follows:\\

    A clear statistical separation is observed between the Schwarzschild geometry and the other two models. The Schwarzschild fit gives a reduced $\chi^{2} = 1.57 (540.26/344)$. In contrast, the JMN-1 model produces a significantly smaller reduced $\chi^{2} = 1.41 (481.63/342)$. Next, comparing the information criteria: the Schwarzschild model gives AIC = 552.26 and BIC = 575.46, while the JMN-1 model gives AIC = 497.63 and BIC = 528.51. The large differences, $\Delta\chi^{2} \approx 58.63$, $\Delta\mathrm{AIC} \approx 55$ and $\Delta\mathrm{BIC} \approx 47$, strongly distinguish Schwarzschild and JMN-1 spacetime. This statistical preference has a clear physical origin. In the Schwarzschild spacetime, the accretion disk is truncated at the ISCO located at $6M$, which suppresses emission from the deepest gravitational potential and leads to a loss in high-energy X-ray flux. In contrast, the JMN-1 naked singularity admits stable circular orbits up to arbitrarily small radii. As a result, the inner disk contributes more efficiently to the high-energy emission, allowing the model to reproduce the observed \textit{NuSTAR} spectrum with significantly better accuracy.\\

    However, when comparing the JMN-1 naked singularity with the Kerr black hole model, the situation changes both qualitatively and quantitatively. The Kerr fit gives a reduced $\chi^{2} \approx 1.41$ (484.02/343)  essentially identical to that of the JMN-1 model. The difference in fit quality between the two models is minimal, with $\Delta\chi^{2} \approx 2.4$. Similarly, the information criteria show only marginal differences: the Kerr model yields AIC = 498.02 and BIC = 524.41, compared to AIC = 497.63 and BIC = 528.51 for the JMN-1 case. These small differences fall well below the thresholds required for a statistically significant preference.\\

    The near degeneracy between the JMN-1 and Kerr models can be connected to their similar effective inner disk structure. A rapidly spinning Kerr black hole allows the accretion disk to extend to radii of order $1$--$2M$, comparable to the inner disk extent permitted in the JMN-1 spacetime. Consequently, both geometries enable the disk to probe similarly deep gravitational potentials, producing comparable reflection signatures and high-energy continuum. In this regime, the X-ray reflection spectrum is primarily sensitive to the depth of the potential well rather than to the presence of an event horizon or to whether the deep disk is enabled by rotation or by a horizonless geometry. Although \texttt{relxill} was developed for Kerr spacetime, we use it here as a common phenomenological framework for the hard energy spectrum when comparing different geometries. Key reflection-related parameters, such as the ionization state of the disk, inclination angle, and iron abundance, are primarily determined by accretion disk properties and are not uniquely tied to a specific spacetime metric. Using the same reflection model for all geometries, therefore, allows us to observe the impact of the inner disk structure and emission extent on the resulting spectra. Since no fully self-consistent relativistic reflection model currently exists for horizonless spacetimes, this approach provides an internally consistent framework for comparing the JMN-1 model with Kerr and Schwarzschild geometries.\\

    The information criterion analysis supports this interpretation. Although the JMN-1 model is slightly favored over Kerr by AIC, the Kerr model is marginally preferred by BIC, reflecting the small differences in parameter counting and data weighting. However, these differences are statistically insignificant, indicating that the current data do not allow a robust distinction between a high-spin Kerr black hole and a JMN-1 naked singularity. In contrast, the Schwarzschild geometry is decisively disfavored by both goodness of fit statistics and information criteria. These results imply that X-ray reflection spectroscopy is effective at distinguishing between accretion disks with different inner truncation radii, but is insufficient, by itself, to uniquely identify the nature of the central compact object when different geometries have similar inner disk radius. In particular, a non-spinning naked singularity spacetime such as JMN-1 can remain observationally indistinguishable from a rapidly spinning Kerr black hole, even though their underlying causal structures are fundamentally different.\\

    Recently, the Event Horizon Telescope (EHT) released high-resolution images of Sgr A*, renewing interest in alternative compact-object models. Their shadow analysis showed that the JMN-1 naked singularity can closely reproduce the shadow of a Schwarzschild black hole, even without an event horizon. This makes current shadow based tests alone insufficient for identifying the true nature of the compact object. In contrast, our results show that X-ray reflection spectroscopy can reveal differences between Schwarzschild and JMN-1. Thus, X-ray spectra can provide a potentially more critical probe of horizonless geometries.\\

    A related form of discrepancy appears in spin measurements of compact objects inferred from different observational probes. For example, black hole spins estimated from X-ray reflection or continuum fitting in X-ray binaries are often found to be high, while the effective spins inferred from gravitational-wave observations of binary black hole mergers are typically lower \cite{fishbach2022apples}. Although these measurements probe very different astrophysical environments and evolutionary channels, they illustrate how observationally inferred parameters can depend strongly on the modeling assumptions and the physical processes involved. In this context, our results highlight that spectral signatures alone may not uniquely determine the underlying spacetime geometry of the compact object, and that additional observational probes may be required to understand this.\\

    Overall, our work highlights the need for more accurate, physically consistent reflection models for horizonless spacetimes. Although our analysis shows that Schwarzschild and JMN-1 can be distinguished through their X-ray spectra, the JMN-1 model still mimics the high-spin Kerr case because both allow the disk to extend deep into the potential well. In other words, when two geometries share similar ISCO locations, current reflection models are not sensitive enough to separate them. This highlights the need for future models that incorporate not only ISCO based effects but also additional geometric features that directly reflect rotation and spacetime structure. As next-generation X-ray missions improve in sensitivity and resolution, such refined modeling will be crucial for distinguishing true black holes from their horizonless mimickers.

\section{Acknowledgment}

VP acknowledges D. Tahelyani and A. Joshi for discussions, and the Council of Scientific and Industrial Research (CSIR, India; Ref: 09/1294(18267)/2024-EMR-I) for financial support.
SB acknowledges financial support by the Fulbright-Nehru Academic \& Professional Excellence Award (Research), sponsored by the U.S. Department of State and the United States-India Educational Foundation (grant number: 3062/F-N APE/2024; program number: G-1-00005).
SB also acknowledges the support of the Department of Atomic Energy, Government of India, under Project Identification No. RTI 4012.
We utilized the archived NuSTAR data provided by the High Energy Astrophysics Science Archive Research Center (HEASARC) online service maintained by the Goddard Space Flight Center. For this work, we used the NuSTAR Data Analysis Software (NuSTARDAS) jointly developed by the ASI Space Science Data Center (SSDC, Italy) and the California Institute of Technology (Caltech, USA).

\nocite{*}  % Include all references from the .bib file
\bibliography{references}  % Use the correct .bib file name

@PREAMBLE{
 "\providecommand{\noopsort}[1]{}" 
 # "\providecommand{\singleletter}[1]{#1}%" 
}

@article{Penrose:1964wq,
    author = "Penrose, Roger",
    title = "{Gravitational collapse and space-time singularities}",
    doi = "10.1103/PhysRevLett.14.57",
    journal = "Phys. Rev. Lett.",
    volume = "14",
    pages = "57--59",
    year = "1965"
}

@article{Joshi:1993zg,
    author = "Joshi, P. S. and Dwivedi, I. H.",
    title = "{Naked singularities in spherically symmetric inhomogeneous Tolman-Bondi dust cloud collapse}",
    eprint = "gr-qc/9303037",
    archivePrefix = "arXiv",
    reportNumber = "TIFR-TAP-9-92",
    doi = "10.1103/PhysRevD.47.5357",
    journal = "Phys. Rev. D",
    volume = "47",
    pages = "5357--5369",
    year = "1993"
}

@article{Deshingkar:1998ge,
    author = "Deshingkar, S. S. and Jhingan, S. and Joshi, P. S.",
    title = "{On the global visibility of singularity in quasispherical collapse}",
    eprint = "gr-qc/9806055",
    archivePrefix = "arXiv",
    doi = "10.1023/A:1018813108516",
    journal = "Gen. Rel. Grav.",
    volume = "30",
    pages = "1477--1499",
    year = "1998"
}

@article{Ortiz:2011jw,
    author = "Ortiz, Nestor and Sarbach, Olivier",
    title = "{Conformal diagrams for the gravitational collapse of a spherical dust cloud}",
    eprint = "1106.2504",
    archivePrefix = "arXiv",
    primaryClass = "gr-qc",
    doi = "10.1088/0264-9381/28/23/235001",
    journal = "Class. Quant. Grav.",
    volume = "28",
    pages = "235001",
    year = "2011"
}

@article{Debnath:2003pq,
    author = "Debnath, Ujjal and Chakraborty, Subenoy and Barrow, John D.",
    title = "{Quasispherical gravitational collapse in any dimension}",
    eprint = "gr-qc/0305075",
    archivePrefix = "arXiv",
    doi = "10.1023/B:GERG.0000010472.10539.46",
    journal = "Gen. Rel. Grav.",
    volume = "36",
    pages = "231--243",
    year = "2004"
}

@article{Giacomazzo:2011cv,
    author = "Giacomazzo, Bruno and Rezzolla, Luciano and Stergioulas, Nikolaos",
    title = "{Collapse of differentially rotating neutron stars and cosmic censorship}",
    eprint = "1105.0122",
    archivePrefix = "arXiv",
    primaryClass = "gr-qc",
    doi = "10.1103/PhysRevD.84.024022",
    journal = "Phys. Rev. D",
    volume = "84",
    pages = "024022",
    year = "2011"
}

@article{Ori:1989ps,
    author = "Ori, Amos and Piran, Tsvi",
    title = "{Naked Singularities and Other Features of Selfsimilar General Relativistic Gravitational Collapse}",
    reportNumber = "HEBREW-5",
    doi = "10.1103/PhysRevD.42.1068",
    journal = "Phys. Rev. D",
    volume = "42",
    pages = "1068--1090",
    year = "1990"
}

@article{Goswami:2004ne,
    author = "Goswami, Rituparno and Joshi, Pankaj S.",
    title = "{Naked singularity formation in scalar field collapse}",
    eprint = "gr-qc/0410144",
    journal = "-",
    archivePrefix = "arXiv",
    month = "10",
    year = "2004"
}

@article{Joshi:2011zm,
    author = "Joshi, Pankaj S. and Malafarina, Daniele and Narayan, Ramesh",
    title = "{Equilibrium configurations from gravitational collapse}",
    eprint = "1106.5438",
    archivePrefix = "arXiv",
    primaryClass = "gr-qc",
    doi = "10.1088/0264-9381/28/23/235018",
    journal = "Class. Quant. Grav.",
    volume = "28",
    pages = "235018",
    year = "2011"
}

@article{Mosani:2020mro,
    author = "Mosani, Karim and Dey, Dipanjan and Joshi, Pankaj S.",
    title = "{Global visibility of a strong curvature singularity in nonmarginally bound dust collapse}",
    eprint = "2003.07092",
    archivePrefix = "arXiv",
    primaryClass = "gr-qc",
    doi = "10.1103/PhysRevD.102.044037",
    journal = "Phys. Rev. D",
    volume = "102",
    number = "4",
    pages = "044037",
    year = "2020"
}

@article{Mosani:2021byw,
    author = "Mosani, Karim and Dey, Dipanjan and Joshi, Pankaj S.",
    title = "{Globally visible singularity in an astrophysical setup}",
    eprint = "2103.07179",
    archivePrefix = "arXiv",
    primaryClass = "gr-qc",
    doi = "10.1093/mnras/stab1186",
    journal = "Mon. Not. Roy. Astron. Soc.",
    volume = "504",
    number = "4",
    pages = "4743--4750",
    year = "2021"
}

@article{Joshi:2013dva,
    author = "Joshi, Pankaj S. and Malafarina, Daniele and Narayan, Ramesh",
    title = "{Distinguishing black holes from naked singularities through their accretion disc properties}",
    eprint = "1304.7331",
    archivePrefix = "arXiv",
    primaryClass = "gr-qc",
    doi = "10.1088/0264-9381/31/1/015002",
    journal = "Class. Quant. Grav.",
    volume = "31",
    pages = "015002",
    year = "2014"
}

@article{Bambhaniya:2019pbr,
    author = "Bambhaniya, Parth and Joshi, Ashok B. and Dey, Dipanjan and Joshi, Pankaj S.",
    title = "{Timelike geodesics in Naked Singularity and Black Hole Spacetimes}",
    eprint = "1908.07171",
    archivePrefix = "arXiv",
    primaryClass = "gr-qc",
    doi = "10.1103/PhysRevD.100.124020",
    journal = "Phys. Rev. D",
    volume = "100",
    number = "12",
    pages = "124020",
    year = "2019"
}

@article{Dey:2020haf,
    author = "Dey, Dipanjan and Shaikh, Rajibul and Joshi, Pankaj S.",
    title = "{Perihelion precession and shadows near black holes and naked singularities}",
    eprint = "2003.06810",
    archivePrefix = "arXiv",
    primaryClass = "gr-qc",
    doi = "10.1103/PhysRevD.102.044042",
    journal = "Phys. Rev. D",
    volume = "102",
    number = "4",
    pages = "044042",
    year = "2020"
}

@article{Shaikh:2019hbm,
  author = {Shaikh, Rajibul and Joshi, Pankaj S.},
  title = {Can we distinguish black holes from naked singularities by the images of their accretion disks?},
  journal = {JCAP},
  year = {2019},
  volume = {2019},
  number = {10},
  pages = {064},
  doi = {10.1088/1475-7516/2019/10/064},
  eprint = {1909.10322},
  archivePrefix = {arXiv},
  primaryClass = {gr-qc}
}

@article{Trivedi:2025ukh,
    author = "Trivedi, Jay Verma and Joshi, Pankaj S. and Gopal-Krishna and Biermann, Peter L.",
    title = "{Astrophysical Black holes: An Explanation for the Galaxy Quenching}",
    eprint = "2501.00899",
    archivePrefix = "arXiv",
    primaryClass = "gr-qc",
    journal = "-",
    month = "1",
    year = "2025"
}

@article{Tahelyani:2022uxw,
    author = "Tahelyani, Divya and Joshi, Ashok B. and Dey, Dipanjan and Joshi, Pankaj S.",
    title = "{Comparing thin accretion disk properties of naked singularities and black holes}",
    eprint = "2205.04055",
    archivePrefix = "arXiv",
    primaryClass = "gr-qc",
    doi = "10.1103/PhysRevD.106.044036",
    journal = "Phys. Rev. D",
    volume = "106",
    number = "4",
    pages = "044036",
    year = "2022"
}

@article{Guo:2020tgv,
    author = "Guo, Jun-Qi and Joshi, Pankaj S. and Narayan, Ramesh and Zhang, Lin",
    title = "{Accretion disks around naked singularities}",
    eprint = "2011.06154",
    archivePrefix = "arXiv",
    primaryClass = "gr-qc",
    doi = "10.1088/1361-6382/abce44",
    journal = "Class. Quant. Grav.",
    volume = "38",
    number = "3",
    pages = "035012",
    year = "2021"
}

@article{Shaikh:2018lcc,
    author = "Shaikh, Rajibul and Kocherlakota, Prashant and Narayan, Ramesh and Joshi, Pankaj S.",
    title = "{Shadows of spherically symmetric black holes and naked singularities}",
    eprint = "1802.08060",
    archivePrefix = "arXiv",
    primaryClass = "astro-ph.HE",
    doi = "10.1093/mnras/sty2624",
    journal = "Mon. Not. Roy. Astron. Soc.",
    volume = "482",
    number = "1",
    pages = "52--64",
    year = "2019"
}

@article{Broderick:2024vjp,
    author = "Broderick, Avery E. and Salehi, Kiana",
    title = "{Cosmic Censorship in Sgr A* and M87*: Observationally Excluding Naked Singularities}",
    eprint = "2406.05181",
    archivePrefix = "arXiv",
    primaryClass = "astro-ph.HE",
    doi = "10.3847/1538-4357/ad90aa",
    journal = "Astrophys. J.",
    volume = "977",
    number = "2",
    pages = "249",
    year = "2024"
}

@article{Chakraborty:2024jma,
    author = "Chakraborty, Chandrachur and Bhattacharyya, Sudip and Joshi, Pankaj S.",
    title = "{Low mass naked singularities from dark core collapse}",
    eprint = "2405.08758",
    archivePrefix = "arXiv",
    primaryClass = "astro-ph.HE",
    doi = "10.1088/1475-7516/2024/07/053",
    journal = "JCAP",
    volume = "07",
    pages = "053",
    year = "2024"
}

@article{Kalsariya:2024qyp,
    author = "Kalsariya, Viraj and Bambhaniya, Parth and Joshi, Pankaj S.",
    title = "{Relativistic time delay analysis of pulsar signals near ultracompact objects}",
    eprint = "2405.01835",
    archivePrefix = "arXiv",
    primaryClass = "gr-qc",
    doi = "10.1103/PhysRevD.110.104026",
    journal = "Phys. Rev. D",
    volume = "110",
    number = "10",
    pages = "104026",
    year = "2024"
}

@article{Joshi:2024gog,
    author = "Joshi, Pankaj S. and Bhattacharyya, Sudip",
    title = "{Primordial naked singularities}",
    eprint = "2401.14431",
    archivePrefix = "arXiv",
    primaryClass = "gr-qc",
    doi = "10.1088/1475-7516/2025/01/034",
    journal = "JCAP",
    volume = "01",
    pages = "034",
    year = "2025"
}

@article{Azreg-Ainou:2023izc,
    author = {Azreg-A{\"\i}nou, Mustapha and Acharya, Kauntey and Joshi, Pankaj S.},
    title = "{Joshi{\textendash}Malafarina{\textendash}Narayan singularity in weak magnetic field}",
    eprint = "2310.08022",
    archivePrefix = "arXiv",
    primaryClass = "gr-qc",
    doi = "10.1140/epjc/s10052-024-12905-4",
    journal = "Eur. Phys. J. C",
    volume = "84",
    number = "5",
    pages = "535",
    year = "2024"
}

@article{Saurabh:2023otl,
    author = "Saurabh, Saurabh and Bambhaniya, Parth and Joshi, Pankaj S.",
    title = "{Imaging ultracompact objects with radiatively inefficient accretion flows}",
    eprint = "2308.14519",
    archivePrefix = "arXiv",
    primaryClass = "astro-ph.HE",
    doi = "10.1051/0004-6361/202347941",
    journal = "Astron. Astrophys.",
    volume = "682",
    pages = "A113",
    year = "2024"
}

@article{Pal:2022cxb,
    author = "Pal, Kunal and Pal, Kuntal and Roy, Pratim and Sarkar, Tapobrata",
    title = "{Regularizing the JNW and JMN naked singularities}",
    eprint = "2206.11764",
    archivePrefix = "arXiv",
    primaryClass = "gr-qc",
    doi = "10.1140/epjc/s10052-023-11558-z",
    journal = "Eur. Phys. J. C",
    volume = "83",
    number = "5",
    pages = "397",
    year = "2023"
}

@article{Acharya:2023vlv,
    author = "Acharya, Kauntey and Bambhaniya, Parth and Joshi, Pankaj S. and Pandey, Kshitij and Patel, Vishva",
    title = "{High energy particle collisions in the vicinity of naked singularity}",
    eprint = "2303.16590",
    archivePrefix = "arXiv",
    primaryClass = "gr-qc",
    doi = "10.1016/j.dark.2025.102101",
    journal = "Phys. Dark Univ.",
    volume = "50",
    pages = "102101",
    year = "2025"
}

@article{Bambhaniya:2022xbz,
    author = "Bambhaniya, Parth and Joshi, Ashok B. and Dey, Dipanjan and Joshi, Pankaj S. and Mazumdar, Arindam and Harada, Tomohiro and Nakao, Ken-ichi",
    title = "{Relativistic orbits of S2 star in the presence of scalar field}",
    eprint = "2209.12610",
    archivePrefix = "arXiv",
    primaryClass = "gr-qc",
    reportNumber = "RUP-22-20, AP-GR-183, NITEP 146",
    doi = "10.1140/epjc/s10052-024-12477-3",
    journal = "Eur. Phys. J. C",
    volume = "84",
    number = "2",
    pages = "124",
    year = "2024"
}

@article{EventHorizonTelescope:2022xqj,
    author = "Akiyama, Kazunori and others",
    collaboration = "Event Horizon Telescope",
    title = "{First Sagittarius A* Event Horizon Telescope Results. VI. Testing the Black Hole Metric}",
    eprint = "2311.09484",
    archivePrefix = "arXiv",
    primaryClass = "astro-ph.HE",
    reportNumber = "FERMILAB-PUB-22-422-PPD",
    doi = "10.3847/2041-8213/ac6756",
    journal = "Astrophys. J. Lett.",
    volume = "930",
    number = "2",
    pages = "L17",
    year = "2022"
}

@article{Liu:2018bfx,
    author = "Liu, Honghui and Zhou, Menglei and Bambi, Cosimo",
    title = "{Distinguishing black holes and naked singularities with iron line spectroscopy}",
    eprint = "1801.00867",
    archivePrefix = "arXiv",
    primaryClass = "gr-qc",
    doi = "10.1088/1475-7516/2018/08/044",
    journal = "JCAP",
    volume = "08",
    pages = "044",
    year = "2018"
}

@inproceedings{Bambi:2021hxv,
    author = "Bambi, Cosimo",
    title = "{Testing General Relativity with black hole X-ray data: recent progress and future developments}",
    booktitle = "{55th Rencontres de Moriond on QCD and High Energy Interactions}",
    eprint = "2103.11365",
    archivePrefix = "arXiv",
    primaryClass = "gr-qc",
    month = "3",
    year = "2021"
}

@book{Bambi:2022iuz,
    editor = "Bambi, Cosimo and Santangelo, Andrea",
    title = "{Handbook of X-ray and Gamma-ray Astrophysics}",
    doi = "10.1007/978-981-16-4544-0",
    isbn = "978-981-16-4544-0",
    publisher = "Springer",
    year = "2024"
}

@article{Dovciak:2003jym,
    author = "Dovciak, M. and Karas, Vladimir and Yaqoob, T.",
    title = "{An Extended scheme for fitting x-ray data with accretion disk spectra in the strong gravity regime}",
    eprint = "astro-ph/0403541",
    archivePrefix = "arXiv",
    doi = "10.1086/421115",
    journal = "Astrophys. J. Suppl.",
    volume = "153",
    pages = "205--221",
    year = "2003"
}

@article{Psaltis:2008bb,
    author = "Psaltis, Dimitrios",
    title = "{Probes and Tests of Strong-Field Gravity with Observations in the Electromagnetic Spectrum}",
    eprint = "0806.1531",
    archivePrefix = "arXiv",
    primaryClass = "astro-ph",
    doi = "10.12942/lrr-2008-9",
    journal = "Living Rev. Rel.",
    volume = "11",
    pages = "9",
    year = "2008"
}

@article{Reynolds:2011sba,
    author = "Reynolds, C. S. and Brenneman, L. W. and Lohfink, A. M. and Trippe, M. L. and Miller, J. M. and Reis, R. C. and Nowak, M. A. and Fabian, A. C.",
    editor = "Petre, Rob and Mitsuda, Kazuhisa and Angelini, Lorella",
    title = "{Probing Relativistic Astrophysics Around SMBHs: The Suzaku AGN Spin Survey}",
    eprint = "1112.0036",
    archivePrefix = "arXiv",
    primaryClass = "astro-ph.HE",
    doi = "10.1063/1.3696170",
    journal = "AIP Conf. Proc.",
    volume = "1427",
    number = "1",
    pages = "157--164",
    year = "2012"
}

@article{Johannsen:2012ng,
    author = "Johannsen, Tim and Psaltis, Dimitrios",
    title = "{Testing the No-Hair Theorem with Observations in the Electromagnetic Spectrum. IV. Relativistically Broadened Iron Lines}",
    eprint = "1202.6069",
    archivePrefix = "arXiv",
    primaryClass = "astro-ph.HE",
    doi = "10.1088/0004-637X/773/1/57",
    journal = "Astrophys. J.",
    volume = "773",
    pages = "57",
    year = "2013"
}

@article{Miller:2012zj,
    author = "Miller, J. M. and Pooley, G. G. and Fabian, A. C. and Nowak, M. A. and Reis, R. C. and Cackett, E. M. and Pottschmidt, K. and Wilms, J.",
    title = "{On the Role of the Accretion Disk in Black Hole Disk-Jet Connections}",
    eprint = "1207.3752",
    archivePrefix = "arXiv",
    primaryClass = "astro-ph.HE",
    doi = "10.1088/0004-637X/757/1/11",
    journal = "Astrophys. J.",
    volume = "757",
    pages = "11",
    year = "2012"
}

@article{Bambi:2012tg,
    author = "Bambi, Cosimo",
    title = "{A code to compute the emission of thin accretion disks in non-Kerr space-times and test the nature of black hole candidates}",
    eprint = "1210.5679",
    archivePrefix = "arXiv",
    primaryClass = "gr-qc",
    doi = "10.1088/0004-637X/761/2/174",
    journal = "Astrophys. J.",
    volume = "761",
    pages = "174",
    year = "2012"
}

@article{Bambi:2013qj,
    author = "Bambi, Cosimo",
    title = "{Testing the space-time geometry around black hole candidates with the available radio and X-ray data}",
    eprint = "1301.0361",
    archivePrefix = "arXiv",
    primaryClass = "gr-qc",
    doi = "10.1080/21672857.2013.11519712",
    journal = "Astron. Rev.",
    volume = "8",
    number = "1",
    pages = "4--39",
    year = "2013"
}

@article{Bambi:2013hza,
    author = "Bambi, Cosimo and Malafarina, Daniele",
    title = "{K$\alpha$ iron line profile from accretion disks around regular and singular exotic compact objects}",
    eprint = "1307.2106",
    archivePrefix = "arXiv",
    primaryClass = "gr-qc",
    doi = "10.1103/PhysRevD.88.064022",
    journal = "Phys. Rev. D",
    volume = "88",
    pages = "064022",
    year = "2013"
}

@article{Jiang:2014loa,
    author = "Jiang, Jiachen and Bambi, Cosimo and Steiner, James F.",
    title = "{Using iron line reverberation and spectroscopy to distinguish Kerr and non-Kerr black holes}",
    eprint = "1406.5677",
    archivePrefix = "arXiv",
    primaryClass = "gr-qc",
    doi = "10.1088/1475-7516/2015/05/025",
    journal = "JCAP",
    volume = "05",
    pages = "025",
    year = "2015"
}

@article{Johannsen:2014arp,
    author = "Johannsen, Tim",
    title = "{X-ray Probes of Black Hole Accretion Disks for Testing the No-Hair Theorem}",
    eprint = "1501.02815",
    archivePrefix = "arXiv",
    primaryClass = "astro-ph.HE",
    doi = "10.1103/PhysRevD.90.064002",
    journal = "Phys. Rev. D",
    volume = "90",
    number = "6",
    pages = "064002",
    year = "2014"
}

@article{Riaz:2022rlx,
    author = "Riaz, Shafqat and Shashank, Swarnim and Roy, Rittick and Abdikamalov, Askar B. and Ayzenberg, Dimitry and Bambi, Cosimo and Zhang, Zuobin and Zhou, Menglei",
    title = "{Testing regular black holes with X-ray and GW data}",
    eprint = "2206.03729",
    archivePrefix = "arXiv",
    primaryClass = "gr-qc",
    doi = "10.1088/1475-7516/2022/10/040",
    journal = "JCAP",
    volume = "10",
    pages = "040",
    year = "2022"
}

@article{Kurmanov:2025uwq,
    author = "Kurmanov, Yergali and Boshkayev, Kuantay and Konysbayev, Talgar and Muccino, Marco and Luongo, Orlando and Urazalina, Ainur and Dalelkhankyzy, Anar and Belissarova, Farida and Alimkulova, Madina",
    title = "{Accretion disk luminosity around rotating naked singularities}",
    eprint = "2503.18028",
    archivePrefix = "arXiv",
    primaryClass = "gr-qc",
    doi = "10.1016/j.dark.2025.101917",
    journal = "Phys. Dark Univ.",
    volume = "48",
    pages = "101917",
    year = "2025"
}

@inproceedings{Novikov:1973kta,
    author = "Novikov, I. D. and Thorne, K. S.",
    title = "{Astrophysics and black holes}",
    booktitle = "{Les Houches Summer School of Theoretical Physics}: {Black Holes}",
    pages = "343--550",
    year = "1973"
}

@article{Page:1974he,
    author = "Page, Don N. and Thorne, Kip S.",
    title = "{Disk-Accretion onto a Black Hole. Time-Averaged Structure of Accretion Disk}",
    doi = "10.1086/152990",
    journal = "Astrophys. J.",
    volume = "191",
    pages = "499--506",
    year = "1974"
}

@article{Tripathi:2020cje,
    author = "Tripathi, Ashutosh and Liu, Honghui and Bambi, Cosimo",
    title = "{Impact of the reflection model on the estimate of the properties of accreting black holes}",
    eprint = "2007.15914",
    archivePrefix = "arXiv",
    primaryClass = "astro-ph.HE",
    doi = "10.1093/mnras/staa2618",
    journal = "Mon. Not. Roy. Astron. Soc.",
    volume = "498",
    number = "3",
    pages = "3565--3577",
    year = "2020"
}

@article{Jiang:2022sqv,
    author = "Jiang, Jiachen and Abdikamalov, Askar B. and Bambi, Cosimo and Reynolds, Christopher S.",
    title = "{Black hole spin measurements based on a thin disc model with finite thickness {\textendash} I. An example study of MCG{\ensuremath{-}}06-30-15}",
    eprint = "2205.06696",
    archivePrefix = "arXiv",
    primaryClass = "astro-ph.HE",
    doi = "10.1093/mnras/stac1369",
    journal = "Mon. Not. Roy. Astron. Soc.",
    volume = "514",
    number = "3",
    pages = "3246--3259",
    year = "2022"
}

@article{Tripathi:2017uif,
    author = "Tripathi, Ashutosh",
    editor = "Nicolini, Piero and Kaminski, Matthias and Mureika, Jonas and Bleicher, Marcus",
    title = "{Testing general relativity with X-ray reflection spectroscopy of MCG-06-30-15}",
    eprint = "1709.10213",
    archivePrefix = "arXiv",
    primaryClass = "astro-ph.HE",
    doi = "10.1088/1742-6596/942/1/012017",
    journal = "J. Phys. Conf. Ser.",
    volume = "942",
    number = "1",
    pages = "012017",
    year = "2017"
}

@article{Kammoun:2017wcq,
    author = "Kammoun, E. S. and Papadakis, I. E.",
    title = "{The nature of X-ray spectral variability in MCG{\textendash}6-30-15}",
    eprint = "1709.01059",
    archivePrefix = "arXiv",
    primaryClass = "astro-ph.HE",
    doi = "10.1093/mnras/stx2181",
    journal = "Mon. Not. Roy. Astron. Soc.",
    volume = "472",
    number = "3",
    pages = "3131--3146",
    year = "2017"
}

@article{Lira:2015wya,
    author = "Lira, P. and Arevalo, P. and Uttley, P. and McHardy, I. M. M. and Videla, L.",
    title = "{Long-term monitoring of the archetype Seyfert galaxy MCG-6-30-15: X-ray, optical and near-IR variability of the corona, disc and torus}",
    eprint = "1508.05928",
    archivePrefix = "arXiv",
    primaryClass = "astro-ph.GA",
    doi = "10.1093/mnras/stv1945",
    journal = "Mon. Not. Roy. Astron. Soc.",
    volume = "454",
    number = "1",
    pages = "368--379",
    year = "2015"
}

@article{Emmanoulopoulos:2011ur,
    author = "Emmanoulopoulos, D. and McHardy, I. M. and Papadakis, I. E.",
    title = "{Negative X-ray reverberation time delays from MCG-6-30-15 and Mrk 766}",
    eprint = "1106.6067",
    archivePrefix = "arXiv",
    primaryClass = "astro-ph.CO",
    doi = "10.1111/j.1745-3933.2011.01106.x",
    journal = "Mon. Not. Roy. Astron. Soc.",
    volume = "416",
    number = "1",
    pages = "L94--L98",
    year = "2011"
}

@article{Reynolds:1995mt,
    author = "Reynolds, C. S. and Fabian, A. C. and Nandra, K. and Inoue, H. and Kunieda, H. and Iwasawa, K.",
    title = "{Asca pv observations of the seyfert 1 galaxy mcg-6-30-15 : rapid variability of the warm absorber}",
    eprint = "astro-ph/9506086",
    archivePrefix = "arXiv",
    reportNumber = "CAP-9506010",
    doi = "10.1093/mnras/277.3.901",
    journal = "Mon. Not. Roy. Astron. Soc.",
    volume = "277",
    pages = "901",
    year = "1995"
}

@article{Iwasawa:1996uh,
    author = "Iwasawa, K. and others",
    title = "{The Variable iron k emission line in MCG-6-30-15}",
    eprint = "astro-ph/9606103",
    archivePrefix = "arXiv",
    doi = "10.1093/mnras/282.3.1038",
    journal = "Mon. Not. Roy. Astron. Soc.",
    volume = "282",
    pages = "1038--1048",
    year = "1996"
}

@article{Orr:1997mf,
    author = "Orr, Astrid and Molendi, S. and Fiore, F. and Grandi, P. and Parmar, A. N. and Owens, Alan",
    title = "{Soft x-ray observations of the complex warm absorber in mcg-6-30-15 with bepposax}",
    eprint = "astro-ph/9706133",
    archivePrefix = "arXiv",
    journal = "Astron. Astrophys.",
    volume = "324",
    pages = "L77",
    year = "1997"
}

@article{Weaver:1998ad,
    author = "Weaver, Kimberly A. and Yaqoob, Tahir",
    title = "{On the evidence for extreme gravity effects in mcg-6-30-15}",
    eprint = "astro-ph/9806045",
    archivePrefix = "arXiv",
    doi = "10.1086/311497",
    journal = "Astrophys. J. Lett.",
    volume = "502",
    pages = "L139",
    year = "1998"
}

@article{Nowak:1999dj,
    author = "Nowak, Michael A. and Chiang, James",
    title = "{Implications of the x-ray variability for the mass of mcg-6-30-15}",
    eprint = "astro-ph/9906371",
    archivePrefix = "arXiv",
    doi = "10.1086/312508",
    journal = "Astrophys. J. Lett.",
    volume = "531",
    pages = "L13",
    year = "2000"
}

@article{Fabian:2002gj,
    author = "Fabian, A. C. and Vaughan, Simon and Nandra, K. and Iwasawa, K. and Ballantyne, D. R. and Lee, J. C. and De Rosa, A. and Turner, A. and Young, A. J.",
    title = "{A Long hard look at MCG-6-30-15 with XMM-Newton}",
    eprint = "astro-ph/0206095",
    archivePrefix = "arXiv",
    doi = "10.1046/j.1365-8711.2002.05740.x",
    journal = "Mon. Not. Roy. Astron. Soc.",
    volume = "335",
    pages = "L1",
    year = "2002"
}

@article{Zycki:2010jd,
    author = "Zycki, P. T. and Ebisawa, K. and Niedzwiecki, A. and Miyakawa, T.",
    title = "{On the light-bending model of X-ray variability of MCG-6-30-15}",
    eprint = "1006.4552",
    archivePrefix = "arXiv",
    primaryClass = "astro-ph.HE",
    doi = "10.1093/pasj/62.5.1185",
    journal = "Publ. Astron. Soc. Jap.",
    volume = "62",
    pages = "1185",
    year = "2010"
}

@article{Noda:2011my,
    author = "Noda, Hirofumi and Makishima, Kazuo and Uehara, Yuuichi and Yamada, Shin'ya and Nakazawa, Kazuhiro",
    title = "{Suzaku Discovery of a Hard Component Varying Independently of the Power-Law Emission in MCG-6-30-15}",
    eprint = "1106.5872",
    archivePrefix = "arXiv",
    primaryClass = "astro-ph.CO",
    doi = "10.1093/pasj/63.2.449",
    journal = "Publ. Astron. Soc. Jap.",
    volume = "63",
    pages = "449",
    year = "2011"
}

@article{Parker:2013cia,
    author = "Parker, M. L. and Marinucci, A. and Brenneman, L. and Fabian, A. C. and Kara, E. and Matt, G. and Walton, D. J.",
    title = "{Principal component analysis of MCG{\textendash}06-30-15 with XMM{\textendash}Newton}",
    eprint = "1310.1945",
    archivePrefix = "arXiv",
    primaryClass = "astro-ph.HE",
    doi = "10.1093/mnras/stt1925",
    journal = "Mon. Not. Roy. Astron. Soc.",
    volume = "437",
    number = "1",
    pages = "721--729",
    year = "2014"
}

@article{Marinucci:2014ita,
    author = "Marinucci, A. and others",
    title = "{The Broadband Spectral Variability of MCG{\textendash}6-30-15 Observed by $NuSTAR$ and $XMM-Newton$}",
    eprint = "1404.3561",
    archivePrefix = "arXiv",
    primaryClass = "astro-ph.HE",
    doi = "10.1088/0004-637X/787/1/83",
    journal = "Astrophys. J.",
    volume = "787",
    pages = "83",
    year = "2014"
}

@article{Gupta:2018soy,
    author = "Gupta, Alok C. and Tripathi, Ashutosh and Wiita, Paul J. and Gu, Minfeng and Bambi, Cosimo and Ho, Luis C.",
    title = "{Possible {\textasciitilde}1 hour quasi-periodic oscillation in narrow-line Seyfert 1 galaxy MCG{\textendash}06{\textendash}30{\textendash}15}",
    eprint = "1808.05112",
    archivePrefix = "arXiv",
    primaryClass = "astro-ph.HE",
    doi = "10.1051/0004-6361/201833629",
    journal = "Astron. Astrophys.",
    volume = "616",
    pages = "L6",
    year = "2018"
}

@article{brenneman2025sharper,
  title={A Sharper View of the X-Ray Spectrum of MCG--6-30-15 with XRISM, XMM-Newton, and NuSTAR},
  author={Brenneman, Laura W and Wilkins, Daniel R and Ogorza{\l}ek, Anna and Rogantini, Daniele and Fabian, Andrew C and Garc{\'\i}a, Javier A and Jur{\'a}{\v{n}}ov{\'a}, Anna and Mizumoto, Misaki and Noda, Hirofumi and Behar, Ehud and others},
  journal={The Astrophysical Journal},
  volume={995},
  number={2},
  pages={200},
  year={2025},
  publisher={The American Astronomical Society}
}

@article{arnaud2003x,
  title={An X-Ray Spectral Fitting Package},
  author={Arnaud, Keith and Dorman, Ben and Gordon, Craig},
  journal={Citeseer (Citeseer)},
  year={2003}
}

@article{harrison2013nuclear,
  title={The nuclear spectroscopic telescope array (NuSTAR) high-energy X-ray mission},
  author={Harrison, Fiona A and Craig, William W and Christensen, Finn E and Hailey, Charles J and Zhang, William W and Boggs, Steven E and Stern, Daniel and Cook, W Rick and Forster, Karl and Giommi, Paolo and others},
  journal={The Astrophysical Journal},
  volume={770},
  number={2},
  pages={103},
  year={2013},
  publisher={The American Astronomical Society}
}

@article{fishbach2022apples,
  title={Apples and oranges: comparing black holes in X-ray binaries and gravitational-wave sources},
  author={Fishbach, Maya and Kalogera, Vicky},
  journal={The Astrophysical Journal Letters},
  volume={929},
  number={2},
  pages={L26},
  year={2022},
  publisher={The American Astronomical Society}
}
\bibliographystyle{elsarticle-harv}  % Apply the bibliography style

\end{document}